\documentclass[aps,prb,twocolumn,longbibliography,amsmath,amssymb,superscriptaddress,bibnotes]{revtex4-2}
\usepackage{amssymb}
\usepackage{amsmath}
\usepackage{bm}
\usepackage[dvips]{graphicx}
\usepackage{dcolumn}
\usepackage{longtable}
\usepackage{color}
\usepackage{bm}
\usepackage[version = 4]{mhchem}
\usepackage[resetlabels,labeled]{multibib}
\newcommand{\TSC}{$T_{\mathrm{c}}$}
\newcommand{\dv}{$\bm{d}$}

\begin{document}
\title{$b$-axis and $c$-axis Knight shift measurements in the superconducting state on ultraclean UTe$_2$ with $T_c$ = 2.1~K}
\author{H. Matsumura}
\email{matsumura.hiroki.75r@st.kyoto-u.ac.jp}
\author{Y. Takahashi}
\author{K. Kinjo}
\thanks{Present address: Institute of Multidisciplinary Research for Advanced Materials, Tohoku University, Sendai, Miyagi 980-8577, Japan}
\author{S. Kitagawa}
\author{K. Ishida}
\email{kishida@scphys.kyoto-u.ac.jp}
\affiliation{Department of Physics, Graduate School of Science, Kyoto University, Kyoto 606-8502, Japan}
\author{Y. Tokunaga}
\author{H. Sakai}
\author{S. Kambe}
\affiliation{Advanced Science Research Center, Japan Atomic Energy Agency, Tokai, Ibaraki 319-1195, Japan}
\author{A. Nakamura}
\author{Y. Shimizu}
\author{Y. Homma}
\author{D. Li}
\author{F. Honda}
\thanks{Present address: Central Institute of Radioisotope Science and Safety, Kyushu University, Fukuoka 819-0395, Japan}
\author{A. Miyake}
\affiliation{Institute for Materials Research, Tohoku University, Oarai, Ibaraki 311-1313, Japan}
\author{D. Aoki}
\affiliation{Institute for Materials Research, Tohoku University, Oarai, Ibaraki 311-1313, Japan}
\affiliation{Universit\'e Grenoble Alpes, CEA, IRIG, PHELIQS, F-38000 Grenoble, France}
\date{\today}

\begin{abstract}
Knight shifts along the $b$ and $c$ axes ($K_b$ and $K_c$) at two crystallographically distinct Te sites were measured down to 70 mK using $^{125}$Te nuclear magnetic resonance (NMR) on an ultraclean \ce{UTe2} single crystal with a superconducting (SC) transition temperature \TSC \ = 2.1~K.
This was carried out to determine the \dv-vector components, which are the order parameter in the spin-triplet pairing.
Although the decrease in $K_b$ and $K_c$ is comparable to the theoretical estimation of the SC diamagnetic shielding effect, it is confirmed, by taking the difference between two Knight shifts at the distinct Te sites, that the spin susceptibility along the $b$ and $c$ axes decreases in the SC state.
Taking into account the large decrease in $K_a$ in the SC state, we conclude that the \dv\ vector has components along all three crystal axes.
\end{abstract}
\maketitle

\section{INTRODUCTION}
The superconductivity in \ce{UTe2} was discovered at the end of 2018\cite{RanScience2019}, and after its discovery, various measurements were carried out to investigate its physical properties of \ce{UTe2}\cite{AokiJPCM2022_UTe2review,LewinRPP2023_UTe2review}. 
\ce{UTe2} crystallizes in the orthorhombic, centrosymmetric structure (space group No. 71, $D_{2h}^{25}$ $Immm$), and the U atoms form a two-leg ladder with legs along the $a$ axis and rungs along the $c$ axis [Fig.~\ref{f1}(a)].
Soon after the discovery, \ce{UTe2} has been considered to be a spin-triplet superconductor, because its normal and superconducting (SC) properties are similar to those of U-based ferromagnetic (FM) superconductors\cite{RanScience2019,AokiJPSJ2019_FMSCreview,TokunagaPRL2015,TokunagaJPSJ2022,TokunagaPRL2023,MineevJPSJ2024}.
However, unlike these FM superconductors, UTe$_{2}$ exhibits superconductivity in the paramagnetic state and does not show any FM ordering even in the high-pressure region, where superconductivity suddenly disappears\cite{BraithwaiteCommunPhys2019, ThomasSciAdv2020, KinjoPRB2022, AjeeshJPSJ2024}.

After the discovery, we have measured the $^{125}$Te-NMR Knight shifts of \ce{UTe2} to investigate the spin susceptibility in the SC state, because we consider that the difference between spin-singlet and spin-triplet superconductivity would appear in the spin susceptibility in the SC state. 
The Knight shift probing the static field at the nuclear site is one of the most reliable methods to measure the spin susceptibility in the SC state.
In the spin-triplet pairing, the decrease in the spin part of the Knight shift indicates the presence of a \dv-vector component.
Here, the \dv~vector, which was introduced to understand the superfluid $^3$He and the order parameter in triplet pairing, is defined to correspond to the direction of zero-spin projection\cite{LeggettRMP1975}.
We measured the Knight shift in the early-stage sample with the SC transition temperature \TSC\ $\sim 1.6$ K (1.6 K sample), in which a large residual electronic term was observed at $T \rightarrow 0$ in the specific-heat measurement, and reported that the Knight shifts along the $b$ and $c$ axes ($K_b$ and $K_c$) slightly decrease\cite{NakamineJPSJ2019, NakaminePRB2021, NakamineJPSJ2021}, but the Knight shift along the $a$ axis $K_a$ is almost invariant in the SC state\cite{FujibayashiJPSJ2022}. 
Taking into account these results, we suggested that the dominant SC state would be $B_{3u}$, which does not have an $a$-axis component in the \dv~vector\cite{FujibayashiJPSJ2022}.
This is consistent with the spin anisotropy in the normal state.    
We also investigated how the \dv~vector is changed in $H \parallel b$\cite{KinjoPRB2023} and under hydrostatic pressure in the 1.6 K sample\cite{KinjoSciAdv2023}.
However, it was found that there is a tiny uranium deficiency in the 1.6 K sample\cite{HagaJPCM2022}. 
Quite recently, we reported a reduction in $K_a$ in the 1.6~K sample with Knight shift measurements using the field tilting method in the $ab$ plane\cite{KitagawaJPSJ2024}. 
In the paper, it was clarified that the previous invariant $K_a$ originates from the non-SC fraction remaining in the sample, since the tiny uranium deficiency induces the non-SC region in the 1.6 K sample, which is the origin of the residual electronic term.

Owing to improving the quality of the \ce{UTe2} sample, we could utilize an ultraclean single-crystalline sample with \TSC\ $\sim 2.1$~K (2.1 K sample), which is regarded as a disorder-free sample from the relationship between \TSC\ and the residual electronic term\cite{AokiJPSJ2024_UTe2cooking}.
We reported a large reduction in $K_a$ in the SC state of the 2.1~K  sample\cite{MatsumuraJPSJ2023}. 
As discussed in the paper, a large Knight shift reduction gives rise to the Pauli-depairing effect if superconductivity were in the spin-singlet pairing.
However, such an effect was not observed at around the Pauli-depairing field $\mu_0 H_{\mathrm{P}} \sim 1.9$ T estimated from the large reduction in $K_a$ in \ce{UTe2}.
Rather, we pointed out that the absence of the Pauli-depairing effect and the large upper critical field $\mu_0 H_{\mathrm{c2}} \sim 12$ T is consistent with the spin-triplet SC state with the alignment of the SC spin by magnetic fields.

It should be noted that the decrease in $K_a$ in the SC state is more than 20 times larger than those in $K_b$ and $K_c$, and that the decrease in $K_b$ and $K_c$ is only $\sim 3$\% of the Knight shift in the normal state\cite{MatsumuraJPSJ2023}. 
In such a small decrease below \TSC, it is crucially important to conclude whether or not the small decrease in $K_b$ and $K_c$ means a reduction in spin susceptibility in the SC state of the ultraclean 2.1 K sample.
This is because the SC diamagnetic effect gives a similar negative shift to the decrease in spin susceptibility.

In this paper, we succeeded in separating the Knight shift decrease due to these two different origins by measuring $K_b$ and $K_c$ at the two crystallographically distinct Te sites.
We show that the decrease in $K_b$ and $K_c$ in the SC state cannot be interpreted solely by the SC diamagnetic effect, but should instead be ascribed to the decrease in spin susceptibility.
The decrease in spin susceptibility along the $b$ and $c$ axes indicates the finite component of the \dv~vector along the $b$ and $c$ axes.

\section{EXPERIMENTAL METHODS }
\subsection{Samples}
We used an ultraclean sample with enriched $^{125}$Te with a size of $3 \times 1 \times 0.5$ mm$^3$, which was used in the previous measurement\cite{MatsumuraJPSJ2023,MatsumuraPRB2025}.
The sample was prepared with the newly developed molten salt flux method\cite{SakaiPRM2022} with natural U and 99.9\% $^{125}$Te-enriched metals for starting elements. 
Single crystals under an optimized growth condition with excess uranium exhibit a sharp SC transition at $T_\mathrm{c}$ = 2.06~K, which was determined with the specific heat and ac susceptibility measurements.
The characterization of the sample was described in a supplemental material of the previous article\cite{MatsumuraJPSJ2023}.
\subsection{NMR measurements}
\begin{figure*}[t]
\begin{center}
\includegraphics[]{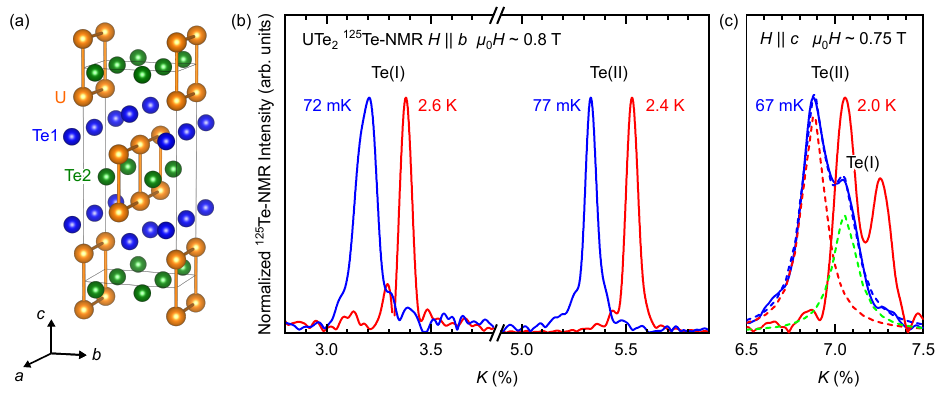}
\end{center}
\caption{(Color online) (a) Crystal structure of \ce{UTe2} drawn by VESTA\cite{vesta}. Te(I) and Te(II) NMR spectra measured at temperatures above \TSC \ and at low temperatures in the SC state in (b) $H \parallel b$ and (c) $H \parallel c$. To determine the peak positions in the double-peak spectrum below \TSC, the spectrum was fitted with the two Lorentian functions, and $K_{c,\mathrm{I}}$ and $K_{c,\mathrm{II}}$ were obtained from the peaks. }
\label{f1}
\end{figure*}

The NMR spectra as a function of frequency were recorded using the Fourier transform of a spin-echo signal observed after a radio-frequency (RF) pulse sequence at a fixed magnetic field. 
The magnetic field was calibrated using a $^{65}$Cu ($^{65}\gamma /2\pi = 12.089$ MHz/T, $K = 0.2385$\%)-NMR signal from a Cu-NMR coil\cite{GCCarter1976}. 
We observed two $^{125}$Te-NMR signals (gyromagnetic ratio $^{125}\gamma/2\pi$ = 13.454 MHz/T, and nuclear spin $I$ = 1/2), when $H$ is applied to the $b$ and $c$ axes, respectively, since there are two crystallographically distinct Te sites in \ce{UTe2}, as shown in Fig. \ref{f1}(a).
Following the previous papers, we call the $^{125}$Te-NMR signal with the smaller [larger] Knight shift value the Te(I) [Te(II)] signal in $H \parallel b$, and this relationship is reversed in $H \parallel c$, as shown in Figs.~\ref{f1}(b) and \ref{f1}(c)\cite{TokunagaJPSJ2019,NakaminePRB2021,FujibayashiJPSJ2023}.
This was confirmed from the angle dependence of the $^{125}$Te-NMR spectrum in the $bc$ plane\cite{NakaminePRB2021}.
Low-temperature NMR measurements down to $\sim 70$ mK were performed using a $^3$He / $^4$He dilution refrigerator with a uniaxis piezoelectric rotator (ANRv51/ULT/RES+, attocube).
The sample was immersed into the $^3$He / $^4$He mixture and rotated in the $ab$ or $ca$ plane with the rotator to apply the magnetic field $H$ precisely parallel to the $b$ or $c$ axis, respectively.
The NMR spectra of the Te(I) and Te(II) signals were recorded with the same NMR pulse condition.
In the SC state, the energy of the RF pulses was reduced to ensure that the NMR results were unchanged by the power of the RF pulses, and a clear shift at both Te signals was observed in $H \parallel b$ and $H \parallel c$, respectively, as shown in Figs.~\ref{f1}(b) and~\ref{f1}(c).      

\section{EXPERIMENTAL RESULTS}
\begin{figure}[tbp]
\vspace{5mm}
\begin{center}
\includegraphics[]{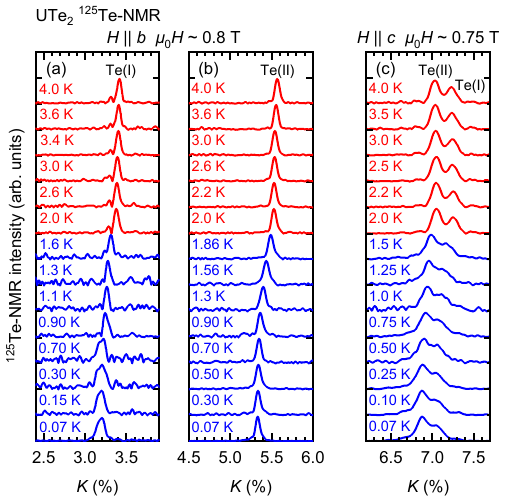}
\end{center}
\caption{(Color online) (a) Te(I) and (b) Te(II)-NMR peaks measured in $\mu_0 H \sim$ 0.8 T along the $b$ axis at various temperatures below 4~K. (c) Two peaks are partially overlapped in $\mu_0 H \sim$ 0.75 T along the $c$ axis, and decrease below \TSC.   }
\label{f2}
\end{figure}
\begin{figure}[tbp]
\vspace{5mm}
\begin{center}
\includegraphics[]{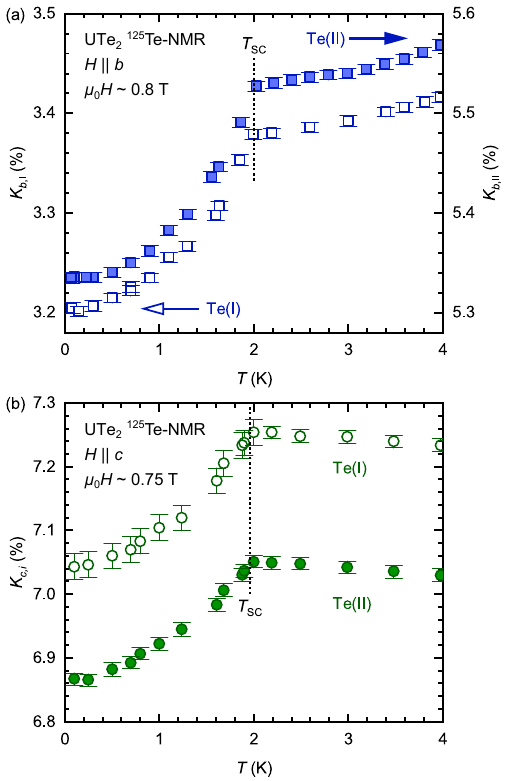}
\end{center}
\caption{(Color online) Temperature dependence of Knight shift measured at the Te(I) and Te(II) signals in (a) $H \parallel b$ and $H \parallel c$ below 4~K.  }
\label{f3}
\end{figure}

\begin{figure}[tbp]
\vspace{5mm}
\begin{center}
\includegraphics[]{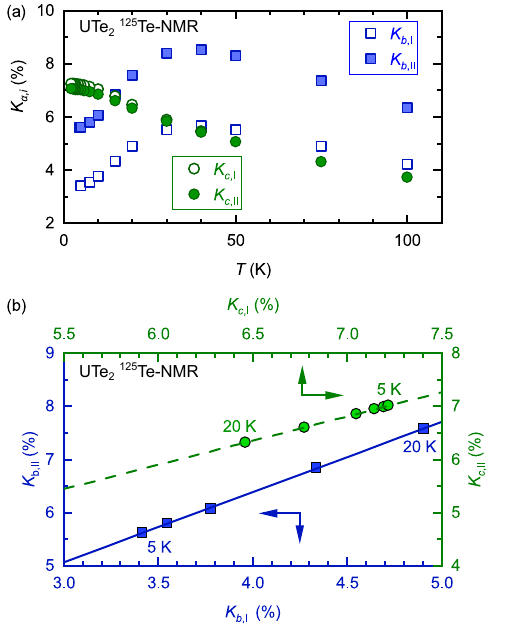}
\end{center}
\caption{(Color online) (a) Temperature dependence of $K_{\alpha, \mathrm{I}}$ and $K_{\alpha, \mathrm{II}}$ ($\alpha$ = $b$ and $c$) below 100 K. (b) The plot of $K_{\alpha, \mathrm{II}}$ against $K_{\alpha, \mathrm{I}}$ ($\alpha$ = $b$ and $c$) below 20~K.}
\label{f4}
\end{figure}

Figures \ref{f2}(a) and \ref{f2}(b) show the Te(I) and Te(II)-NMR spectra measured in $\mu_0 H \sim$ 0.8 T along the $b$ axis, and Fig.~\ref{f2}(c) shows the Te(II) and Te(I)-NMR spectra in $\mu_0 H \sim$ 0.75 T along the $c$ axis at various temperatures below 4~K.
When $H$ is applied to the $b$ axis, the Te(I) and Te(II)-NMR spectra are well separated, but when $H$ is applied to the $c$ axis, two spectra are partially overlapped, and the separation between two peaks becomes unclear in the SC state due to a larger decrease in the Te(I) Knight shift than in Te(II) and the spectrum broadening.
Thus, $K_c$ in the SC state was determined by the two-peak Lorentzian fit, as shown in Fig.~\ref{f1}(c).
Figure \ref{f3}(a) [\ref{f3}(b)] shows the temperature dependence of the Knight shift of the Te(I) and Te(II) signals measured in $\mu_0 H \sim$ 0.8 [0.75] T along the $b$ [$c$] axis below 4~K. 
The $K_b$ of both sites gradually decreases in the normal state below $\sim 35$~K, which is called $T_{\chi_{\mathrm{max}}}$, and the overall temperature dependence in the normal state below 100~K is shown in Fig.~\ref{f4}~(a).
$K_b$ and $K_c$ of both Te(I) and Te(II) signals sharply decrease below \TSC, and are almost saturated at low temperatures.
The magnitude of the decrease in $K_b$ and $K_c$ in the SC state of the 2.1 K sample is almost double that of the 1.6 K sample.
This result is consistent with the specific-heat measurements; while nearly half of the electronic term in the normal state remains at $T \rightarrow 0$ in the 1.6 K sample, the residual electronic term is almost zero in the 2.1 K sample.
The improvement of the superconductivity in the 2.1 K sample was also verified from the present spin-susceptibility measurement in the SC state.

Next, we discuss the temperature dependence of the spin susceptibility in the SC state, not affected by the SC diamagnetic effect.
In order to eliminate the SC diamagnetic effect in the decrease in $K_b$ and $K_c$, we performed the following analyses.
In general, the total Knight shift along the $\alpha$ axis ($\alpha$ = $b$ and $c$) at the Te($i$) ($i$= I or II) signal $ K_{\alpha, i}$ in the SC state is given by  
\begin{align*}
    K_{\alpha, i}(T, H) = K_{\alpha,i}^{\mathrm{spin}}(T) + K_{\alpha, i}^{\mathrm{orb}} + K_{\alpha}^{\mathrm{dia}}(T, H)
\end{align*}
$K_{\alpha,i}^{\mathrm{orb}}$ is the so-called ``orbital'' shift, which results from the field-induced orbital electron magnetism and is considered to be constant at low temperatures. 
$K_{\alpha}^{\mathrm{dia}}(T, H)$ is the shift originating from the SC diamagnetism, which results from the Meissner shielding effect. 
This gives the fractional change in the average internal field in the sample, and is reasonably assumed to be working at both signals in the same manner, and thus site independent.
$K_{\alpha,i}^{\mathrm{spin}}(T)$ is the spin part of the Knight shift, which is proportional to the spin susceptibility $\chi^{\mathrm{spin}}$, and is expressed as $K_{\alpha, i}^{\mathrm{spin}}(T) = A_{\alpha, i} \chi^{\mathrm{spin}}(T)$. 
Here, $A_{\alpha, i}$ is the hyperfine coupling constant along the $\alpha$ axis at the Te($i$) signal.
In most cases, to derive the change in $K_{\alpha, i}^{\mathrm{spin}}(T)$, the magnitude of $K_{\alpha}^{\mathrm{dia}}(T, H)$ was estimated with a theoretical equation\cite{deGennes} with the SC parameters reported experimentally, but it is generally quite hard to know the exact value and the temperature dependence of $K_{\alpha}^{\mathrm{dia}}(T, H)$.  
In \ce{UTe2}, $\chi^{\mathrm{spin}}(T)$ can be derived from the site dependence of $K_{\alpha, i}(T)$ ($i$ = I and II) ; 
\begin{align*}
    K_{\alpha, \mathrm{II}}(T) - K_{\alpha, \mathrm{I}}(T) = (A_{\alpha, \mathrm{II}} - A_{\alpha, \mathrm{I}})\chi^{\mathrm{spin}}(T) +const. ,
\end{align*}
when the Te(I) and Te(II) Knight shifts are scaled to the bulk susceptibility, and several groups reported a good scaling\cite{TokunagaJPSJ2019, AmbikaPRB2022, KinjoPRB2022}.
It is noted the $K_{\alpha}^{\mathrm{dia}}(T, H)$ is canceled out in this method.      

\begin{figure}[tbp]
\vspace{5mm}
\begin{center}
\includegraphics[]{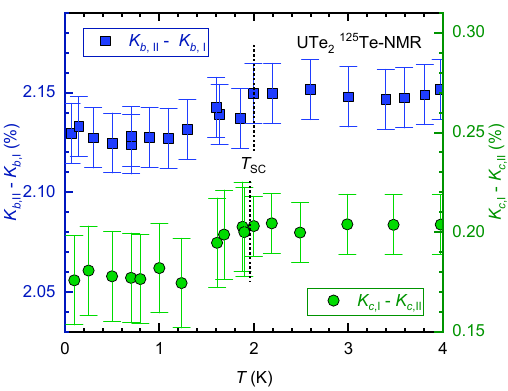}
\end{center}
\caption{(Color online) Temperature dependence of $K_{b, \mathrm{II}} - K_{b, \mathrm{I}}$ and $K_{c, \mathrm{I}} - K_{c, \mathrm{II}}$.}
\label{f5}
\end{figure}

Figure \ref{f5} shows $K_{b, \mathrm{II}} - K_{b, \mathrm{I}}$ for $H \parallel b$ and $K_{c, \mathrm{I}} - K_{b, \mathrm{II}}$ for $H \parallel c$.
It is shown that two quantities are almost constant above \TSC \ but clearly decrease at \TSC, indicative of the decrease in the spin susceptibility along the $b$ and $c$ axis in the SC state.
As mentioned above, we reported a large reduction of $K_a(T)$ that indicates the decrease in the spin susceptibility along the $a$ axis in the SC state\cite{MatsumuraJPSJ2023}.
Thus, from the Knight shift measurements in three axis directions, it is concluded that the \dv~vector has components along all three axes.

\begin{figure}[t]
\begin{center}
\includegraphics[]{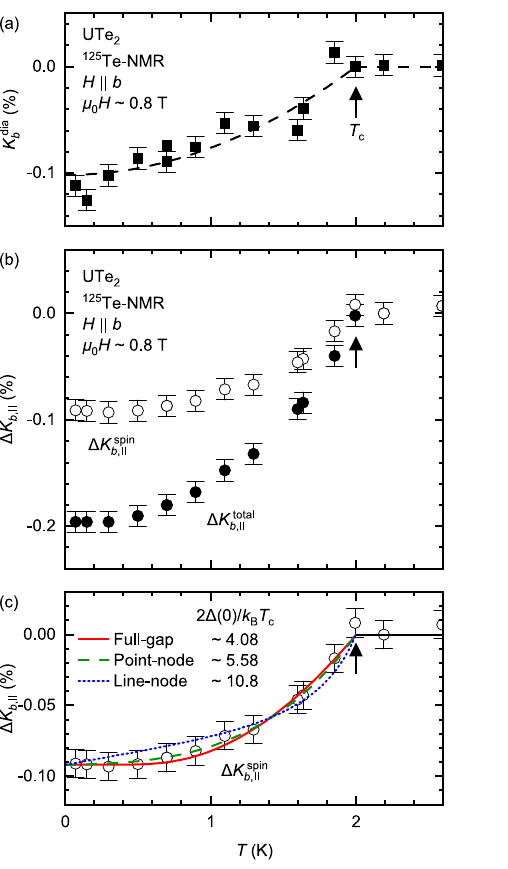}
\end{center}
\caption{
(a) Temperature dependence of $K_b^{\mathrm{dia}}$ evaluated using eq.~(1) based on experimental values.
The fitting curve is shown as a dashed line.
(b) Temperature dependence of $\Delta K_{b,\mathrm{II}}^{\mathrm{total}}$ and $\Delta K_{b,\mathrm{II}}^{\mathrm{spin}}$ evaluated with subtraction of $K_b^{\mathrm{dia}}$.
(c) Fitting curves obtained using various SC models for the temperature dependence of $\Delta K_{b,\mathrm{II}}^{\mathrm{spin}}$ plotted together with $\Delta K_{b,\mathrm{II}}^{\mathrm{spin}}$.
The arrows show the SC transition temperature.
}
\label{f6}
\end{figure}

Next, we try to separate the decrease in $K_{\alpha, i}^{\mathrm{spin}}(T)$ and site independent $K_{\alpha}^{\mathrm{dia}}(T)$ in the SC state with the experimental values. 
The Knight shift decrease $\Delta K_{\alpha, i}^{\mathrm{total}}$ in the SC state of the Te($i$) signal is evaluated from the subtraction $K_{\alpha, i}$ just above \TSC \ ($K_{\alpha,i}^{\mathrm{N}}$), $\Delta K_{\alpha, i}^{\mathrm{total}} \equiv K_{\alpha, i} - K_{\alpha, i}^{\mathrm{N}}$.        
$\Delta K_{\alpha, \mathrm{I}}^{\mathrm{total}}$ and $\Delta K_{\alpha, \mathrm{II}}^{\mathrm{total}}$ are expressed as,
\begin{align*}
\Delta K_{\alpha, \mathrm{I}}^{\mathrm{total}}(T) & =  \Delta K_{\alpha, \mathrm{I}}^{\mathrm{spin}}(T) + K_{\alpha}^{\mathrm{dia}}(T), \hspace{2mm} {\mathrm{and}} \\
\Delta K_{\alpha, \mathrm{II}}^{\mathrm{total}}(T) & =  \Delta K_{\alpha, \mathrm{II}}^{\mathrm{spin}}(T) + K_{\alpha}^{\mathrm{dia}}(T), \hspace{3mm} 
\end{align*}
respectively.
It is noted that $K_{\alpha}^{\mathrm{dia}}$ can be estimated if we can know the relation between $K_{\alpha, \mathrm{I}}^{\mathrm{spin}}$ and $K_{\alpha, \mathrm{II}}^{\mathrm{spin}}$, experimentally.  
As shown in Fig.~\ref{f4}(b), we plot $K_{\alpha, \mathrm{II}}$ against $K_{\alpha,\mathrm{I}}$ as $T$ is an implicit parameter.
The plot was done below 20~K, where the heavy-Fermion (HF) state was stabilized in \ce{UTe2}. 
This is because it was reported that the hyperfine coupling constant was often changed at around the temperature where the $f$-electron character is changed from the localized to itinerant regime in HF compounds\cite{CurroPRB2004}.
Actually, the change of the hyperfine coupling constant in $H \parallel b$ was reported below $T_{\chi_{\mathrm{max}}}$ in \ce{UTe2}\cite{AmbikaPRB2022}.  
From the good linear relation between $K_{b,\mathrm{I}}$ and $K_{b,\mathrm{II}}$, we obtained $\Gamma_{b} \equiv K_{b,\mathrm{II}}/K_{b,\mathrm{I}} = 1.32 \pm 0.01$, and if this relation is assumed to hold in the SC state, $K_b^{\mathrm{dia}}$ is evaluated as 
\begin{align}
K_{b}^{\mathrm{dia}}(T) = [\Gamma_{b} \Delta K_{b,\mathrm{I}}^{\mathrm{total}}(T) - \Delta K_{b, \mathrm{II}}^{\mathrm{total}}(T)] / (\Gamma_b - 1).
\label{eq1}
\end{align}
Using eq.~(\ref{eq1}) with $\Gamma_b$, we evaluate the temperature dependence of $K_b^{\mathrm{dia}}$, as shown in Fig. \ref{f6}(a).
The decrease in $K_b^{\mathrm{dia}}$ below \TSC \ is not so much steeper than that in the experimentally observed $K_b({\mathrm{I}})$ or $K_b({\mathrm{II}})$. 
This is because $K_{\mathrm{dia}}$ is related to the supercurrent density $\mbox{\boldmath{$j$}}$, which is proportional to the superfluid density $n_s$ or $H_{\mathrm{c1}}$ in type II superconductors.
We fit the evaluated $K_{\mathrm{dia}}$ to the temperature dependence of $K_b^{\mathrm{dia}}(0) [1 - (T/T_{\mathrm{c}})^2]$ with $K_b^{\mathrm{dia}}(0) = -0.10$\%. 
The evaluated value of $K_b^{\mathrm{dia}}$(0) is slightly larger than our previously estimation ($K_b^{\mathrm{dia}} \sim -0.06$\%) from the theoretical equation and experimental values of $H_{\mathrm{c1}}$ reported by Ishihara {\it et al.}\cite{IshiharaPRR2023}.
Thus, the decrease of the spin part of the Knight shift $\Delta K_{b, \mathrm{II}}^{\mathrm{spin}}(T)$ can be known from the subtraction of the dashed curve of $K_{b}^{\mathrm{dia}}(T)$ from $\Delta K_{b, i}^{\mathrm{total}}(T)$, as shown in Fig.~\ref{f6}(b). 

It is noteworthy that the temperature dependence of the spin susceptibility evaluated from the above procedure is in good agreement with that evaluated from $K_{b, \mathrm{II}} - K_{b, \mathrm{I}}$ in Fig.~\ref{f5}.
This indicates the validity of the above analyses and estimation.
On the other hand, the same analyses was performed in the $c$-axis Knight shift, but the reasonable value of $K_c^{\mathrm{dia}}$ was not evaluated.
This is mainly because $\Gamma_{c} \equiv K_{c,\mathrm{II}}/K_{c,\mathrm{I}} = 0.91 \pm 0.01$, which means that the Knight shift difference between two signals is so small, which gives a large ambiguity for the estimation of $K_c^{\mathrm{dia}}$, as $\Gamma_{c} -1$ is the denominator in the eq.~(\ref{eq1}).
Another issue we noticed was the uncertainty in determining of $K_{c, \mathrm{I}}$ through the fitting process, as the two-peak structure became unclear in the SC state. 
It is possible that the decrease in $K_{c, \mathrm{I}}$ may have been overestimated due to this fitting uncertainty.
However, the decrease in the interval between the Te(I) and Te(II) peaks in $H \parallel c$ indicates the decrease of the spin susceptibility in the SC state, thus the previous conclusion that $\chi^{\mathrm{spin}}$ decreases below \TSC \ for $H \parallel c$ remains unchanged.  

Next, we discuss the temperature dependence of the spin susceptibility below \TSC\ on the basis of $\Delta K_{b,\mathrm{II}}^{\mathrm{spin}}(T)$.
The spin susceptibility evaluated from $K_{\alpha,\mathrm{II}} - K_{\alpha,\mathrm{I}}$ and $\Delta K_{b,\mathrm{II}}^{\mathrm{spin}}(T)$ remains nearly constant below 1~K.
To investigate the SC gap structure, the low-temperature behavior of $\Delta K_{b,\mathrm{II}}^{\mathrm{spin}}(T)$ was compared with several SC models featuring different gap structures, as shown in Fig.~\ref{f6}~(c).
Although the line-node model can be ruled out, the current data cannot distinguish between the full-gap or point-node scenarios if an SC gap larger than the standard BCS value [$2\Delta(0)/k_{\mathrm{B}} T_\mathrm{c} = 3.5$] is assumed.
This result does not contradict with both the full-gap behavior suggested by thermal-conductivity measurements\cite{SuetsuguSciAdv2024} nor the point-node behavior reported for the 2.1~K sample\cite{IshiharaNatCommun2023,Lee_condmat,HayesPRX2025,HarunaJPSJ2024}.
However, the nearly constant temperature dependence of $\Delta K_{b,\mathrm{II}}^{\mathrm{spin}}(T)$ below 1~K does not exclude the possibility of a finite residual density of states (DOS) near the Fermi energy $E_{\mathrm{F}}$, which also produces an almost temperature-independent behavior in this low-temperature region\cite{IshidaJPSJ1993}.
Consequently, to determine whether such a residual DOS is present and to further clarify whether the SC state is fully gapped or has point nodes, low-temperature measurements of the spin-lattice relaxation rate $1/T_1$, which is highly sensitive to the residual DOS at $E_{\mathrm{F}}$, are crucial.

The main result of the present measurements is that the spin susceptibility in all three axes unambiguously decreases in the SC state, although the decrease in the Knight shift along the $b$ and $c$-axis is very small.
This indicates that the \dv~vector has components along all three axes, which seemingly suggests the $A_{\mathrm{u}}$ state in the odd-parity $p$-wave for a point group with $D_{\mathrm{2h}}$\cite{IshizukaPRL2019}.
In addition, the large reduction in the quantized $a$ axis is also in good agreement with the $A_{\mathrm{u}}$ state\cite{HiranumaJPSJ2021}.  
The SC gap of the $A_{\mathrm{u}}$ state is a nodeless full gap, which seems to be consistent with the present result, as discussed above. 
We consider that the SC gap in \ce{UTe2} would be highly anisotropic even though the SC state is classified as an $A_{\mathrm{u}}$ state, since the electronic properties in the normal state are highly anisotropic.
This situation is quite different from the superfluid $^3$He. 
Alternatively, the experimental fact of the \dv~vector with all three axis components might be interpreted with the triplet pairing with the higher order term, the $f$-wave, as in the case of UPt$_3$\cite{OhmiPRL1993,MachidaJPSJ1995,SaulsJLTP1994,SaulsAdvPhys1994}.
In this case, all $f$-wave triplet SC states are expressed by the \dv~vector with all three axis components. 
The $f$-wave triplet SC state with point-nodes seems to be also consistent with the present results.

In conclusion, from the measurement of the Knight shift of two $^{125}$Te-NMR signals, we conclude that the spin susceptibility along the $b$ and $c$ axes decreases in the SC state.
Taking the large reduction of $K_a$ into account, we conclude that the \dv~vector has components along all three axes, which gives an important constraint for the SC pairing state.
In addition, the finite component of the \dv~vector along each axis gives rise to the spin degrees of freedom.
It is important to investigate how the \dv~vector is changed under external conditions, such as the magnetic field, hydrostatic pressure, and uniaxial strain, which is now in progress.

\section*{Acknowledgments}
The authors would like to thank J. Ishizuka, Y. Yanase, K. Machida, S. Fujimoto, Y. Kasahara, Y. Matsuda, V. P. Mineev, Y. Maeno, S. Yonezawa, J-P. Brison, G. Knebel, W. Knafo, and J. Flouquet for their valuable inputs in our discussions. 
This work was supported by Grants-in-Aid for Scientific Research (KAKENHI Grant No. JP20KK0061, No. JP20H00130, No. JP21K18600, No. JP22H04933, No. JP22H01168, No. JP23H01124, No. JP23K19022, No. JP23K22439 and No. JP23K25821) from the Japan Society for the Promotion of Science, by JST SPRING(Grant No. JPMJSP2110) from Japan Science and Technology Agency, by research support funding from The Kyoto University Foundation, by ISHIZUE 2024 of Kyoto University Research Development Program, by Murata Science and Education Foundation, and by the JGC-S Scholarship Foundation.
In addition, liquid helium is supplied from the Low Temperature and Materials Sciences Division, Agency for Health, Safety and Environment, Kyoto University. 

\vspace{5mm}

\appendix
\section*{Appendix: The calculation of spin susceptibility in the SC state }
In order to discuss the SC gap structure, we compared the temperature dependence of $\Delta K_{b,\mathrm{II}}^{\mathrm{spin}}$ with spin susceptibility in the SC state calculated with three different SC gap models: full gap, point-node gap, and line-node gap.
Temperature dependence of spin susceptibility $\chi(T)$ \cite{YosidaFunc} is calculated with 
\begin{align*}
    \chi(T) = \int \frac{\mathrm{d}\Omega_{k}}{4\pi} \int_0^\infty \mathrm{d}\xi_{k} \frac{1}{2 k_{\mathrm{B}} T} \mathrm{sech}^2\left(\frac{\sqrt{\xi_{k}^2 + \Delta_{k}^{2}(T)}}{2 k_{\mathrm{B}} T}\right),
\end{align*}
where $\xi_k$ is the single-particle energy relative to the Fermi energy, $\Delta_k(T)$ is the momentum-dependent energy gap, and $\Omega_k$ is the solid angle.
The square of $\Delta_k(T)$ is given by
\begin{align*}
    &\Delta^2(T), \hspace{2mm} & &\text{for\ full\ gap},\\
    &\Delta^2(T) (\hat{k}_x^2 + \hat{k}_y^2), \hspace{2mm} & &\text{for\ point-node\ gap},\\
    &\Delta^2(T) \hat{k}_z^2, \hspace{2mm} & &\text{for\ line-node\ gap},
\end{align*}
where $\Delta(T)$ is the temperature-dependent gap obtained by solving the gap equation\cite{TachikiJMMM1985}, and $\hat{k}_x$, $\hat{k}_y$, and $\hat{k}_z$ are the $x$-, $y$-, and $z$-components of the unit wave vector, respectively.
The temperature dependence of $\Delta(T)$ normalized by its zero-temperature value $\Delta(0)$ is shown in Fig.~7.
To compare with the experimental data, we fitted the calculated temperature dependence of the spin susceptibility by fixing the total reduction to 0.092\%.
In this fitting, $2\Delta(0)/k_{\mathrm{B}} T_\mathrm{c}$ was used as a free parameter, and the resulting values for each gap model are shown in Fig.~6~(c).
\begin{figure}[b]
\vspace{5mm}
\begin{center}
\includegraphics[]{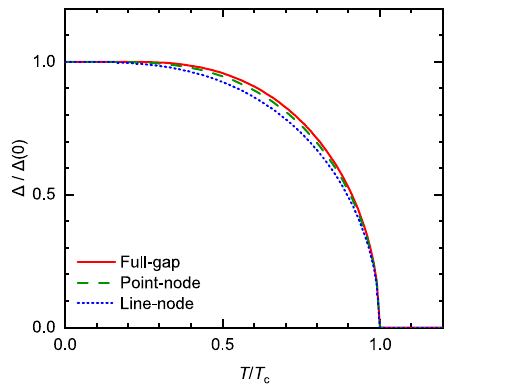}
\end{center}
\caption{
(Color online) Temperature dependence of normalized SC gap $\Delta/\Delta(0)$.
}
\label{f_app}
\end{figure}

\bibliography{UTe2}

\end{document}